\documentclass[fleqn,10pt]{wlscirep}
\usepackage[utf8]{inputenc}
\usepackage[T1]{fontenc}
\usepackage[normalem]{ulem}
\usepackage{multirow}
\usepackage{lineno}
\usepackage{tabularx}
\usepackage{float}
\usepackage{bm}
\usepackage{graphicx}
\usepackage{amsmath}
\usepackage{mathtools}
\usepackage{braket}
\usepackage{color}
\usepackage{mathrsfs}
\usepackage[export]{adjustbox}
\usepackage{bbold}
\usepackage{xspace}

\title{All-optical control of skyrmion configuration in CrI$_3$ monolayer}

\author[1,*]{M. Kazemi}
\author[2,3]{A. Kudlis}
\author[1,4]{P. F. Bessarab}
\author[1,2]{I. A. Shelykh}
\affil[1]{Science Institute, University of Iceland, Dunhagi-3, IS-107 Reykjav\'ik, Iceland}
\affil[2]{Abrikosov Center for Theoretical Physics, MIPT, Dolgoprudnyi, Moscow Region 141701, Russia}
\affil[3]{Russian Quantum Center, Skolkovo, Moscow, 121205, Russia}
\affil[4]{Department of Physics and Electrical Engineering, Linnaeus University, SE-39231 Kalmar, Sweden}

\affil[*]{mak99@hi.is}

\begin{abstract}
The potential for manipulating characteristics of skyrmions in a CrI$_3$ monolayer using circularly polarised light is explored.
The effective skyrmion-light interaction is mediated by bright excitons whose magnetization is selectively influenced by the polarization of  photons. The light-induced skyrmion dynamics is illustrated by the dependencies of the skyrmion size and the skyrmion lifetime on the intensity and polarization of the incident light pulse. Two-dimensional magnets hosting excitons thus represent a promising platform for the control of topological magnetic structures by light.
\end{abstract}
\begin{document}

\flushbottom
\maketitle

\thispagestyle{empty}

\section*{Introduction}
Magnetic materials play a tremendous role in various applications such as magnetic memory where writing of data is associated with change in magnetization. New technological frontiers, such as terahertz speed and high energy efficiency of magnetization switching, can be reached by use of functional materials whose magnetic properties can be all-optically controlled~\cite{du2016weak,matsukura2015control}. Members of the family of 2D magnetic materials known as chromium trihalides (CrX$_3$, where X = I, Br, and Cl), can be considered as prime candidates for this role. Specifically, the optical control of magnetism in monolayers of chromium triiodide (CrI$_3$) was theoretically predicted~\cite{PhysRevB.104.L020412} and experimentally demonstrated~\cite{zhang2022all}. CrI$_3$, an Ising-type ferromagnet, possesses exceptionally high excitonic binding energies and oscillator strengths~\cite{wu2019physical}, making this material distinguished
among similar systems. 

In Ref.~\cite{kudlis2023all} the authors developed a microscopic theory of all-optical resonant control of magnetization in excitonic materials and demonstrated reorientation of lattice magnetisation by incident pulses of light with different polarisations and certain values of light parameters. The phenomenon was elucidated through the transfer of spin angular momentum from the electric field to the excitons of a sample, ultimately acting as an effective magnetic field on the 
lattice magnetisation. However, any analysis of the behaviour of spatially inhomogeneous magnetic structures such as magnetic skyrmions, was not presented and possibility of controlling their properties using excitons remains unexplored. On the other hand, skyrmions themselves are highly appealing for theoretical and experimental studies, owing to their potential utility as units of information storage~\cite{nagaosa2013topological,fert2013skyrmions}.

The formation of stable skyrmions occurs as a result of the interplay between the magnetic exchange, Dzyaloshinskii–Moriya (DM) interaction, magnetic anisotropy, and the Zeeman interaction~\cite{bogdanov1994,bocdanov1994properties,bogdanov1999,rohart2013skyrmion,bacani2019measure,wang2018theory}, while the parameters of the skyrmions such as their size and shape can be controlled by changing the interactions strength. 
However, there remains a 
gap in the development of the theory for all-optical manipulation of skyrmions.

\begin{figure}
    \centering
    \includegraphics[width = 0.50\linewidth]{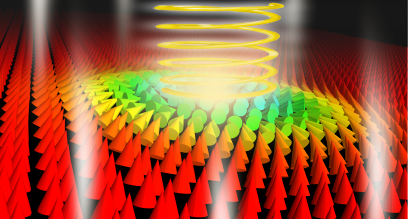}
    \caption{A sketch of all-optical control of a skyrmion in a CrI$_3$ monolayer.
    A light pulse with a certain frequency, polarization and time envelope is incident on the monolayer plane whose lattice magnetization is shown with the arrows. The color codes the out-of-plane projection of the magnetization. Under the influence of increasing exciton magnetization, the spins at the boundary of the skyrmion change their direction, effectively leading to either increase or decrease in the skyrmion size.}

    \label{fig:structure}
\end{figure}

The goal of the present study is to explore the phenomenon of all-optical control over the size of magnetic skyrmions in materials hosting bright excitons. The potential for such manipulation is particularly appealing due to the prospect that achieving smaller radius skyrmions could result in more energy-efficient applications~\cite{wu2021size}. 
We demonstrate that by applying light pulses of varying circular polarisation and intensity, one can control the radius and profile of magnetic N\'eel skyrmions in CrI$_3$ ferromagnetic monolayers (see Fig.~\ref{fig:structure} for the schematic representation of the setup). Additionally, we report light-induced skyrmion collapse, and examine the corresponding skyrmion lifetime as a function of pulse intensity.

The structure of the article is as follows. Following the introduction, we describe the formalism utilised in this study, which is based on coupled dynamics of excitons and lattice magnetization of the system. The results and analysis of numerical simulations of the skyrmion dynamics under light pulses of varying intensities and polarizations are presented in section result and discussion. The final section provides concluding remarks.

\section*{The model and the equations of  motion}
Here, we provide a concise overview of the theoretical framework for all-optical resonant control of magnetization in excitonic materials, but comprehensive details of the theory can be found in Ref.~\cite{kudlis2023all}.

The total energy of the system is given by the following equation:
\begin{equation}
    E=E_{\text{m}}+E_{\text{exc}}+E_{\text{s}},
\end{equation}
which comprises contributions from the lattice magnetization, the excitons, and the interaction between these subsystems. In the formalism, the lattice magnetization 
is modelled as an array of classical vectors 
localized at the hexagonal lattice sites of chromium atoms.
The lattice magnetization is characterized by the following energy: 
\begin{align}
E_{\textup{m}}=-J\sum_{<i,j>}\boldsymbol{m}_i\cdot\boldsymbol{m}_j-D\sum_{<i,j>}\boldsymbol{d}_{ij} \cdot \left[\boldsymbol{m}_i\times\boldsymbol{m}_j\right] 
- K \sum_i\left(\boldsymbol{m}_i\cdot\boldsymbol{e}_z\right)^2-\mu\sum_i \boldsymbol{B}\cdot \boldsymbol{m}_i.
\label{eq:mag_hamiltonian}
\end{align}
Here, the first and the second terms correspond to the Heisenberg exchange and DM %Dzyalashinsky-Maria 
interactions, the third term %is responsible for 
describes the uniaxial anisotropy along the normal %axis perpendicular 
to the plane of the sample, and the last term corresponds to the interaction with an external magnetic field $\boldsymbol{B}$ applied perpendicular to the system's plane. The orientation of magnetic moment of $i$th Cr atom is along the unit vector %The vector 
$\boldsymbol{m}$, and its magnitude $\mu$ is 3 Bohr magnetons. The pairwise interactions are taken only between the nearest neighbors. 
The direction of the DM vector is along the unit vector $\boldsymbol{d}_{ij} = \boldsymbol{r}_{ij}\times \boldsymbol{e}_z/|\boldsymbol{r}_{ij}|$, %is the DMI unit vector 
with $\boldsymbol{e}_z$ and $\boldsymbol{r}_{ij}$ being the unit normal to the monolayer plane and the vector pointing from site $i$ to site $j$, respectively. 
The parameter values used in the present calculations are taken from Ref.~\cite{ghosh2020comment}: $J=2.53$~meV, $D=1.2$~meV, $K=0.153$~meV. For our calculations, we used a simulation domain with the size of 
$N_{\textup{c}}=30 \times 30=900$ unit cells, each of which consists of two Cr atoms.

The 
excitonic subsystem 
is described 
by the following Hamiltonian expressed in terms of the 
creation $\hat{X}^{\dagger}_{n\textbf{q}}$ and annihilation $\hat{X}^{}_{n\textbf{q}}$ operators for excitons:
\begin{align}
\label{eqn:ex_ham}
\hat{H}_{\text{exc}}=\sum_{\textbf{q}n}E_{n\textbf{q}}\hat{X}^{\dagger}_{n\textbf{q}}\hat{X}^{}_{n\textbf{q}} +\boldsymbol{E}_{\pm}\sum_{n}\boldsymbol{D}_{n\textbf{q}=0}\hat{X}^{\dagger}_{0\textbf{q}=0}\hat{X}_{n\textbf{k}=0}+h.c.,     
\end{align}
where $E_{n\textbf{q}}$ 
is the energy of the exciton with the in-plane momentum $\textbf{q}$ and band index $n$, the vector $\boldsymbol{E}{\pm}(t)=\mathrm{Re}[(E_0,\mp\textup{i}E_0,0) \allowbreak \mathrm{exp}(-\textup{i}\omega t)]h(t)$ describes the classical electric field in the light pulse with either right- or left-circular polarization, 
vector $D_{n\textbf{q}=0}$ 
is the dipole moment of the direct-gap optical transition, and
function $h(t)$ describes the time envelope of the light pulse. The effects of radiative and nonradiative scattering are taken into account via the finite inverse relaxation time $\delta$ of the excited exciton states. This involves multiplying the corresponding components of the exciton wave function by a factor $\sim \exp{\left(-\delta \Delta t \right)}$ after each time step $\Delta t$, while preserving the norm of the wave function. Here we use an estimate $\delta = 80$~meV
from Ref.~\cite{Wu_2019}.

The interaction of the excitons with the lattice magnetization is described by the following Hamiltonian:
\begin{align}
   \hat{H}_{\text{s}}=-g\mu\! \! \!\sum_{\substack{\textbf{q}\textbf{q}'nn'}}\!\!\!\left(\boldsymbol{m}_{\textbf{q}-\textbf{q}'}-\boldsymbol{m}^{\text{g}}_{\textbf{q}-\textbf{q}'}\right)\boldsymbol{M}^{\textbf{q}\textbf{q}'}_{nn'}\hat{X}^{\dagger}_{n\textbf{q}}\hat{X}^{}_{n'\textbf{q}'},
\end{align}
where the Fourier transform of the lattice magnetization with respect to the collinear ground state is enclosed in the parentheses. The parameters $E_{n\textbf{q}}$, $D_{n\textbf{q}=0}$, and $\boldsymbol{M}^{\textbf{q}\textbf{q}'}_{nn'}$ were determined using the Density Functional Theory (DFT) calculations using the GPAW
software package~\cite{Enkovaara_2010,Mortensen_2005}. The details of the DFT calculations are provided in our previous work~\cite{kudlis2023all}. Unfortunately, the interaction constant $g$ cannot be computed using standard DFT methods, necessitating its phenomenological definition. In line with our previous study~\cite{kudlis2023all}, we chose 
a value slightly below the Heisenberg interaction constant, setting $g=2.3$ meV.

Note that our analysis excludes exciton-exciton interaction. Therefore, the finite number of excitons in the system is taken into account by introducing the parameter
$n_{\textup{exc}}$ proportional to the number of unit cells, chosen based on considerations of carrier population density (in our case, $n_{\text{exc}}=450$). 
The energy of the interaction between the excitons and the lattice magnetization becomes:
\begin{align}
E_{\text{s}}\approx n_{\textup{exc}}\langle\Psi_{\text{exc}}|\hat{H}_{\text{s}}|\Psi_{\text{exc}}\rangle=-g\mu \sum\limits_{i}\boldsymbol{\sigma}_i(\boldsymbol{m}_i-\boldsymbol{m}_i^{\textup{g}}).
\label{eqn:lat_ham}
\end{align}
Here, $\boldsymbol{m}^{\textup{g}}\equiv \boldsymbol{e}_z$ is the unit vector along the ground-state magnetization, and the exciton spin vector associated with the $i$th unit cell $\boldsymbol{\sigma}_{i}$ is defined via the following equation: 
\begin{align}
    \boldsymbol{\sigma}_i=n_{\text{exc}}\sum_{\textbf{q}\textbf{q}'}e^{2\pi\textup{i}(\boldsymbol{r}_i(\boldsymbol{q}'-\boldsymbol{q}))/N_{\textup{c}}}\sum_{nn'}C_{n}^{\textbf{q}*} \boldsymbol{M}^{\textbf{q}\textbf{q}'}_{nn'}C_{n'}^{\textbf{q}'},\label{eq:sigma}
\end{align}
where $\boldsymbol{r}_i$ is the position of $i^{th}$ unit cell and $C_{n}^{\textbf{q}}$ are the expansion coefficients of the exciton wave function $\Psi_{\text{exc}}$:
\begin{equation}
\Psi_{\text{exc}}=\sum_{n\textbf{q}}C_{n}^{\textbf{q}} \hat{X}^{\dagger}_{n\textbf{q}} |0\rangle.
\end{equation}
\begin{figure}
    \centering
     \includegraphics[width = 0.70\linewidth]{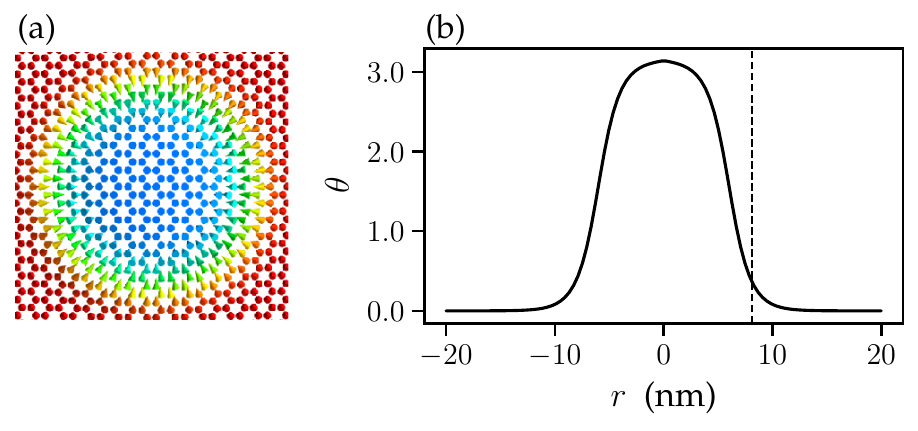}
    \caption{Equilibrium distribution of the magnetization (a) and corresponding magnetization profile (b) for an isolated skyrmion in the CrI$_3$ monolayer at $B = 0.37$ T in the absence of the light pulse. In (a), the color codes the out-of-plane projection of the magnetization. In (b), the dashed vertical line marks the skyrmion radius calculated according to Eq.~\eqref{radius}.}
    \label{fig:theta_profile}
\end{figure}
The exciton dynamics is governed by the following equation:
\begin{align}
\label{eq:c}
i\hbar\frac{\partial C_{n}^{\textbf{q}}}{\partial t}=\sum_{n'\textbf{q}'} H_{nn'}^{\textbf{q}\textbf{q}'}(t) C_{n'}^{\textbf{q}'}(t),    
\end{align}
where
$H_{nn'}^{\textbf{q}\textbf{q}'}$ are the matrix elements of the 
exciton Hamiltonian $\hat{H}=\hat{H}_{\textup{exc}}+\hat{H}_{\textup{s}}$ in the basis 
$\hat{X}^{\dagger}_{n\textbf{q}}\ket{0}$. 

The lattice spin dynamics is described by the Landau–Lifshitz–Gilbert equation:
\begin{align}
    \label{eq:llg}
    \dfrac{d\boldsymbol{m}_i}{dt}=-\gamma \boldsymbol{m}_i\times \boldsymbol{B}^{\text{eff}}_{i}+\eta\left(\boldsymbol{m}_i\times\dfrac{d\boldsymbol{m_i}}{dt}\right),
\end{align}
where $\gamma$ is the gyromagnetic ratio, $\eta = 0.1$ is the dimensionless damping parameter, and $\boldsymbol{B}_i^{\textup{eff}}$ is the effective magnetic field on site $i$:
\begin{align}
\label{eq:beff}
    \boldsymbol{B}^{\text{eff}}_{i}=g\boldsymbol{\sigma}_i - \frac{1}{\mu}\frac{\partial E_m}{\partial\boldsymbol{m}_i}. 
\end{align}
Equation (\ref{eq:llg}) is integrated numerically using the semi-implicit solver~\cite{mentink2010stable}, where the exciton spin $\boldsymbol{\sigma}_i$ is updated every time step using Eqs. (\ref{eq:sigma}) and (\ref{eq:c}).

In this work, we focus on light-driven dynamics of a single isolated skyrmion in the CrI$_3$ monolayer (see Fig. \ref{fig:theta_profile}). Specifically, we study the time evolution of the skyrmion radius, here defined according to Bogdanov and Hubert~\cite{bocdanov1994properties}:
\begin{align}
    R = r_0 - \theta(r_0)\left(\frac{d\theta}{dr}\right)_{r_0}^{-1},\label{radius}
\end{align}
where $\theta (r)$ is the polar angle of the lattice magnetization as a function of the distance to the skyrmion centre and $r_0$ is the steepest slope point. %
\begin{figure}
    \centering
    \includegraphics[width = 0.60\linewidth]{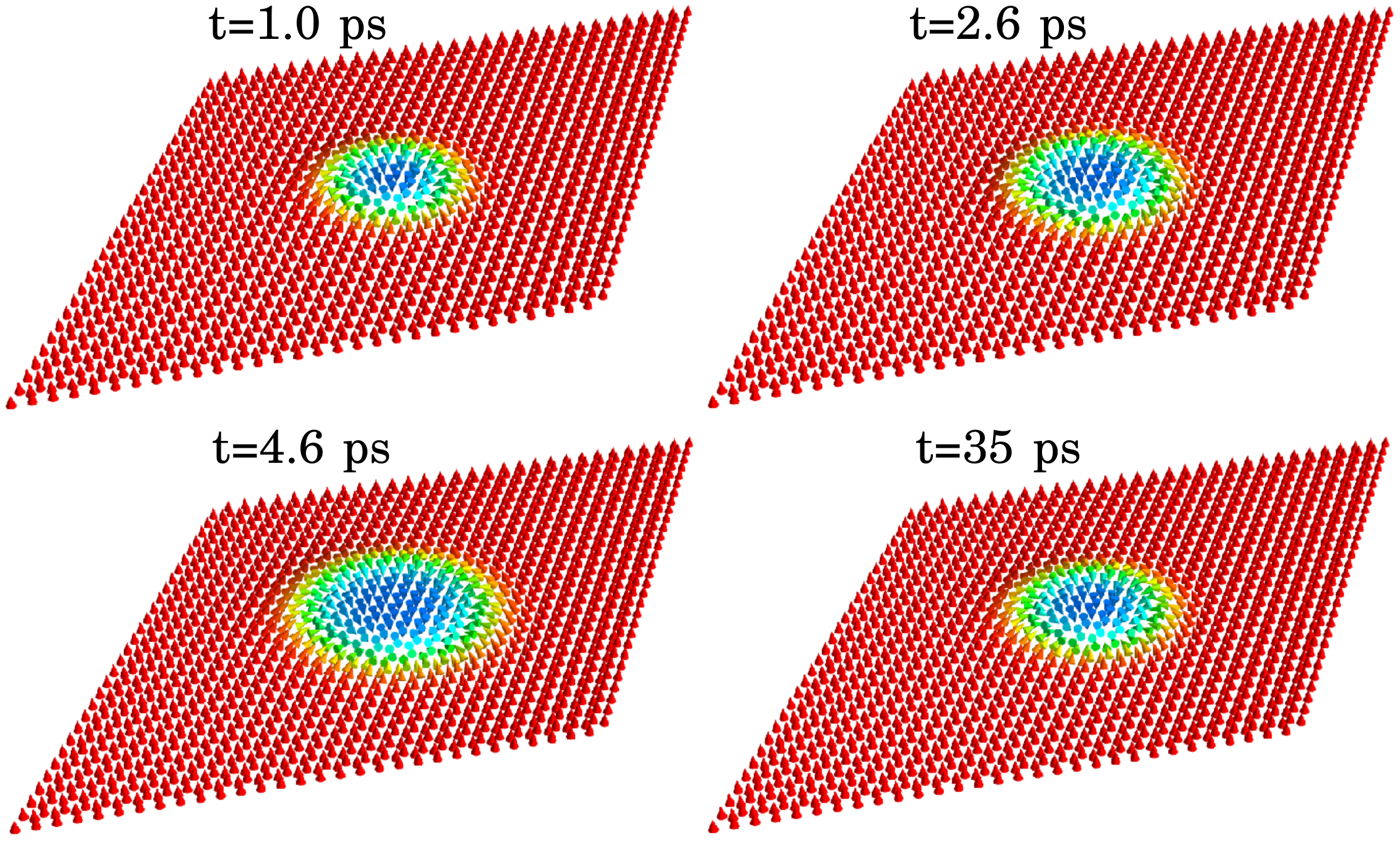}
    \caption{Snapshots of calculated skyrmion dynamics induced by the light pulse with left-handed polarization and fluence $F = 1.028$ mJ/cm$^2$. The labels indicate corresponding instants of time. The color codes the out-of-plane projection of the lattice magnetization. The equilibrium size of the skyrmion, $4.92~\textup{nm}$, is achieved in the external magnetic field of $0.78~\textup{T}$.}
    \label{fig:configtime_small}
\end{figure}
\begin{figure}[!h]
  \centering
    \includegraphics[width=.5\linewidth]{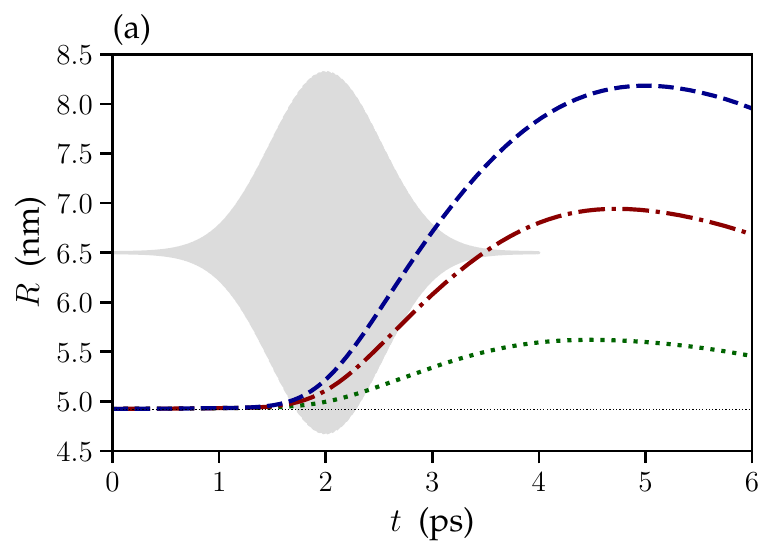}\hfill
    \includegraphics[width=.5\linewidth]{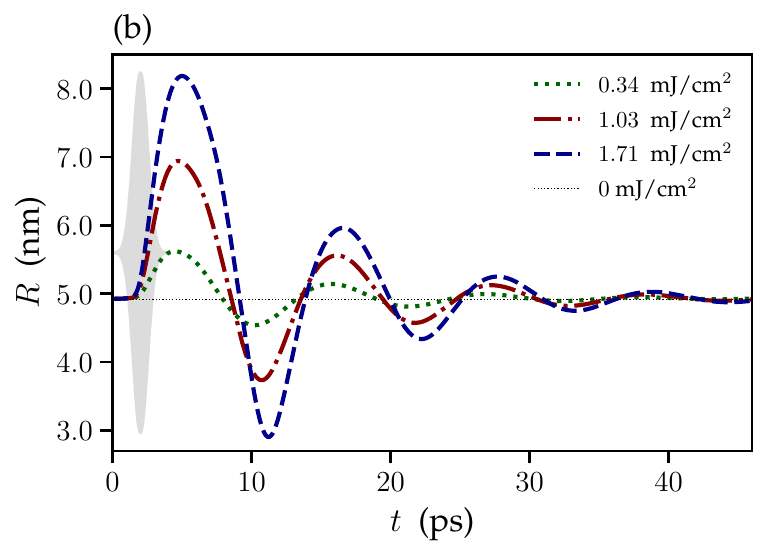}
  \caption{Calculated skyrmion radius as a function of time for several values of the fluence of the applied left-circularly polarized light pulse, as indicated in the legend. The equilibrium size of the skyrmion, 4.92 nm, corresponding to the external magnetic field of 0.78 T is indicated with the dotted horizontal line. Panel (a) corresponds to the beginning of the dynamical process, while panel (b) shows the evolution of the skyrmion radius in a broader range of time. The gray-shaded area shows the time-envelope of the applied light pulse.}
   \label{fig:r_small_as_time}
\end{figure}
\begin{figure}
    \centering
    \includegraphics[width = 0.60\linewidth]{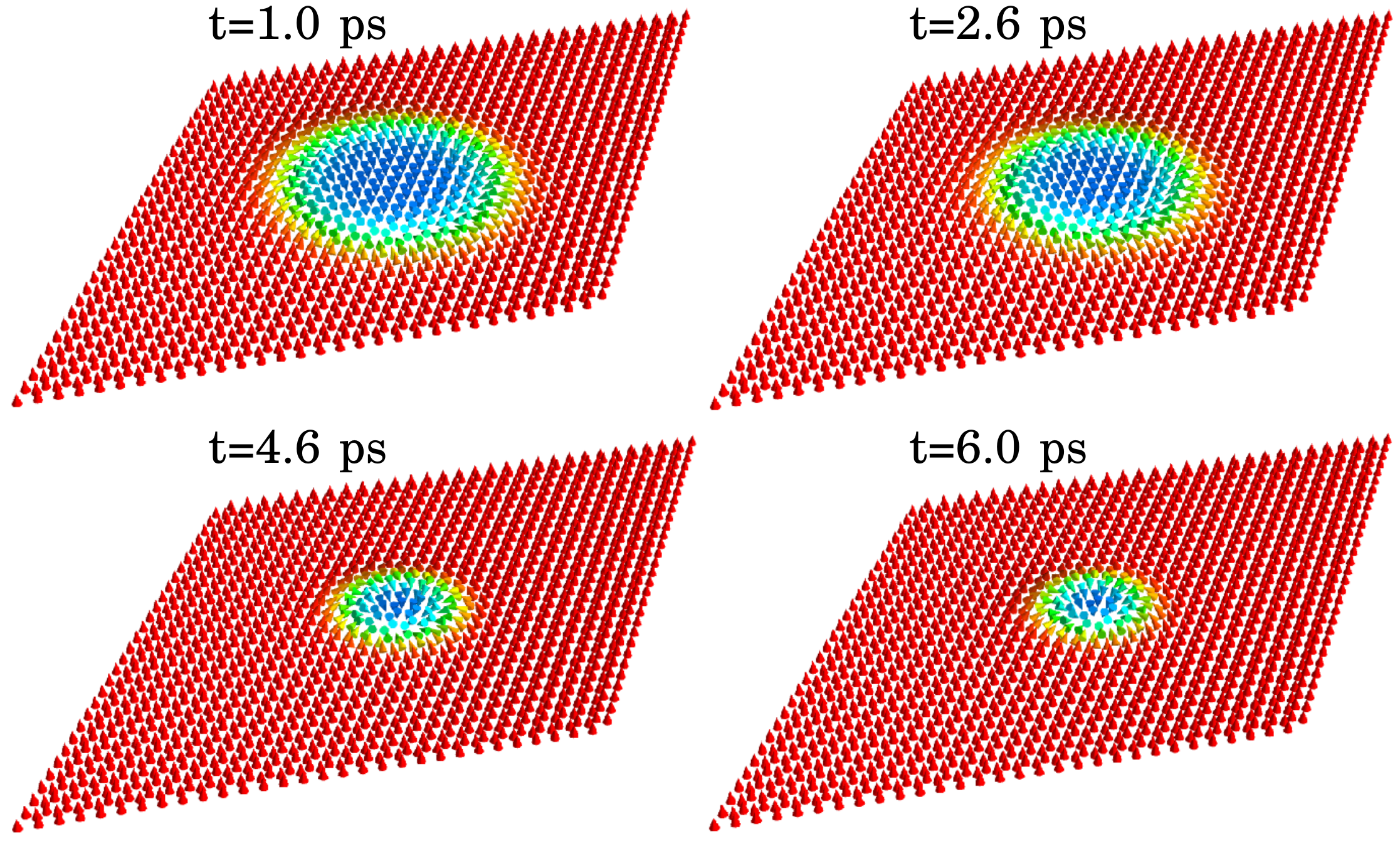}
    \caption{Snapshots of calculated skyrmion dynamics induced by the light pulse with right-handed polarization and fluence $F = 0.342$ mJ/cm$^2$. The labels indicate corresponding instants of time. The color codes the out-of-plane projection of the lattice magnetization. The equilibrium size of the skyrmion, $8.1~\textup{nm}$, is achieved in the external magnetic field of $0.37~\textup{T}$.}
    \label{fig:configtime_large}
\end{figure}
The system is initially prepared at the energy-minimum state corresponding to a single isolated skyrmion whose size can be controlled by adjusting the magnitude of the external magnetic field~\cite{bocdanov1994properties}. At $t=0$, the dynamics of the system is induced by application of the spatially uniform laser pulse characterized by the 
time envelope
$h(t) =\alpha_1 \theta(t_f - |2t -t_f|)\exp{[-\alpha_2 \left((t-t_f/2)/t_f\right)^2]}$ with $t_\text{f}$  being the pulse duration. Dimensionless parameters $\alpha_1=1.94$ and $\alpha_2=30.2$ influence both the shape of the pulse profile and the total pulse fluence $F=E_0^{2}c\varepsilon_0\int_{0}^{t_f}h(t)^2dt/2$. In our calculations, the duration of the pulse $t_f$ is chosen to be $4~\textup{ps}$ and its central frequency $\omega$ is $1.94$ eV, which lies below the direct bandgap.

\section*{Results and discussion}

\begin{figure}[!b]
    \centering
    \includegraphics[width = 0.50\linewidth]{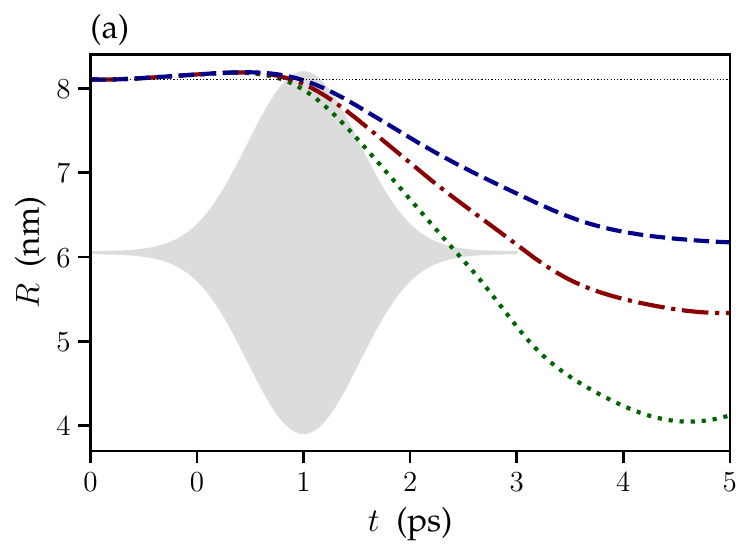}\hfill
    \includegraphics[width = 0.50\linewidth]{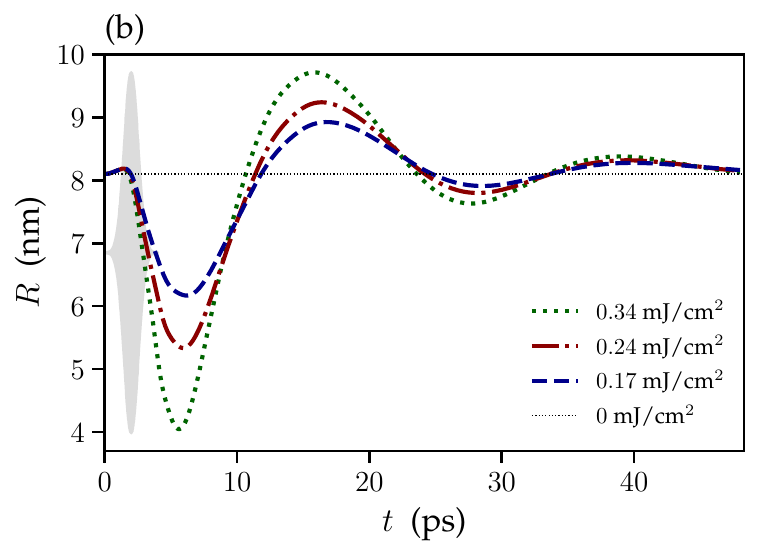}
    \caption{Calculated skyrmion radius as a function of time for several values of the fluence of the applied right-circularly polarized light pulse, as indicated in the legend. The equilibrium size of the skyrmion, 8.1 nm, corresponding to the external magnetic field of 0.37 T is indicated with the dotted horizontal line. Panel (a) corresponds to the beginning of the dynamical process, while panel (b) shows the evolution of the skyrmion radius in a broader range of time. The gray-shaded area shows the time-envelope of the applied light pulse.}
    \label{fig:r_large_as_time}
\end{figure}

Figure~\ref{fig:configtime_small} shows snapshots of the calculated skyrmion configuration at several instants of time during and after the application of the left-circularly polarized light pulse with the fluence $F=1.028$ mJ/cm$^2$.  The light polarization ensures that the magnetization of the induced excitons is along the skyrmion core, which results in the skyrmion growth during the application of the light pulse. As exciton scattering is present in the system, the excitonic magnetization wanes after the application of the pulse, resulting in that the skyrmion relaxes to its equilibrium size of $4.92~\textup{nm}$ corresponding to the static magnetic field of $0.78~\textup{T}$.

Figure~\ref{fig:r_small_as_time} shows the time dependence of the skyrmion radius $R$ corresponding to several values of the fluence of the left-circularly polarized light. As expected, larger fluence results in a faster growth and larger maximum size of the skyrmion. After the application of the pulse, the dynamics is not immediately frozen. Instead, the skyrmion radius demonstrates damped oscillations, eventually converging on the equilibrium value.

Controlled shrinking of the skyrmion is demonstrated in Fig.~\ref{fig:configtime_large} showing the snapshots of the magnetization dynamics induced by the right-circularly polarized light pulse with the fluence $F=0.342$ mJ/cm$^2$. The equilibrium skyrmion radius, $R=8.1~\textup{nm}$, is achieved by application of a static magnetic field of $B = 0.37$ T. In contrast to the case of left-circular polarization, the pulse with the right-circular polarization induces excitons with the magnetization opposite to the skyrmion core. As a result, the skyrmion size decreases. The details of this process can be seen in Fig.~\ref{fig:r_large_as_time} showing the time-dependencies of the skyrmion radius for several values of the pulse fluence. The changes in the skyrmion size are more pronounced for larger fluence values. Interestingly, the skyrmion shape is also affected by the light pulse, which manifests itself in a slight increase in the skyrmion size at the beginning of the dynamical process. Similar to the case of the opposite polarization, termination of the pulse is followed by damped oscillations of the skyrmion radius around the equilibrium value.  

\begin{figure}
    \centering
     \includegraphics[width = 0.60\linewidth]{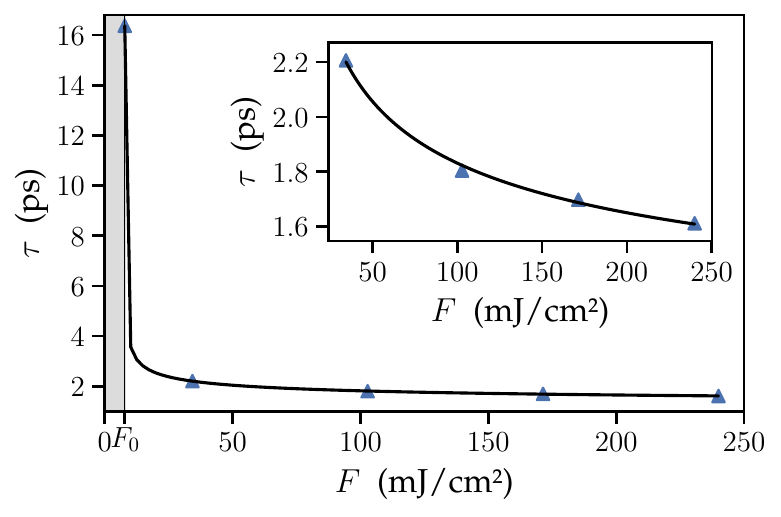}
    \caption{Calculated skyrmion lifetime as a function of the fluence of the incident light pulse with right-circular polarization. The equilibrium skyrmion radius, $8.1~\textup{nm}$, corresponds to a static magnetic field of $0.37$ T. The gray-shaded area marks the fluence magnitudes below the critical value $F_0= 7.84$ mJ/cm$^2$ for which the skyrmion collapse is not possible.}
    \label{fig:lifetime}
\end{figure}

The possibility to control the skyrmion size signifies that the skyrmion can also be destroyed by a light pulse with large enough fluence. To analyze the skyrmion collapse, we apply the right-circularly polarized light of various fluences to the skyrmion with the equilibrium radius of 8.1 nm. In our calculations, we define the skyrmion collaspe as a moment when the skyrmion size has dropped by a factor of 10 compared to its equilibrium level. The corresponding instant of time is referred to as the skyrmion lifetime $\tau$. The calculated skyrmion lifetime as a function of the flueince of the right-circularly polarized light pulse is shown in Fig.~\ref{fig:lifetime}. The dependence is characterized by the critical value of the fluence below which the collapse of the skyrmion is not possible. For the pulse fluence above the critical value, the skyrmion lifetime is finite and decreases monotonically with increasing $F$.

\section*{Conclusion} 
Our study demonstrates the potential for manipulating the size of skyrmions in monolayers of chromium triiodide, without the need for an alternating magnetic field, solely through the use of light pulses with specific polarization and intensity. This achievement holds significant implications for modern spintronics. By employing the combined formalism of atomistic spin dynamics and coupled exciton wave function equations, we demonstrated the possibility to control such important parameters of skyrmions as their radia and lifetimes.

\section*{Acknowledgments}
We are grateful to Y. Zhumagulov for the DFT data for CrI$_3$, as well as to H. Schrautzer for the support with the Spinaker software package for atomistic spin dynamics calculations.  This work is supported by the Ministry of Science and Higher Education of the Russian Federation (Goszadaniye, project No. FSMG-2023-0011), the Icelandic Research Fund (Grants No. 2410537 and No. 217750), the University of Iceland Research Fund (Grant No. 15673), the Swedish Research Council (Grant No. 2020-05110), and the Crafoord Foundation (Grant No. 20231063).

\end{document}